\documentclass{article}

\usepackage{PRIMEarxiv}

\usepackage[utf8]{inputenc} 
\usepackage[T1]{fontenc}    
\usepackage{hyperref}       
\usepackage{url}            
\usepackage{booktabs}       
\usepackage{amsfonts}       
\usepackage{nicefrac}       
\usepackage{microtype}      
\usepackage{lipsum}
\usepackage{fancyhdr}       
\usepackage{graphicx}       
\graphicspath{{media/}}     
\usepackage{svg}            
\usepackage{tabularx}       
\usepackage{placeins}

\title{Optimizing Tensor Contraction Paths: A Greedy Algorithm Approach With Improved Cost Functions}

\author{
  Sheela Orgler \\
  \texttt{sheela.orgler@gmail.com} \\
   \And
  Mark Blacher \\
  \texttt{mark.blacher@uni-jena.de} \\
}

\begin{document}
\maketitle

\begin{abstract}
Finding efficient tensor contraction paths is essential for a wide range of problems, including model counting, quantum circuits, graph problems, and language models.
There exist several approaches to find efficient paths, such as the greedy and random greedy algorithm by Optimized Einsum (opt\_einsum), 
and the greedy algorithm and hypergraph partitioning approach employed in cotengra.
However, these algorithms require a lot of computational time and resources to find efficient contraction paths.
In this paper, we introduce a novel approach based on the greedy algorithm by opt\_einsum that computes efficient contraction paths in less time.
Moreover, with our approach, we are even able to compute paths for large problems where modern algorithms fail.
\end{abstract}


\section{Introduction}
Tensor contractions are central to solving problems in various areas of research such as model counting, quantum circuits~\cite{reinforcement-learning}, graph problems, and machine learning~\cite{hypergraph-partitioning,many-body}.
Finding a contraction order that minimizes the computational cost is crucial.
This can be explained by looking at the computation of the product of a sequence of matrices $A$, $B$, $C$.
The product can be evaluated in two ways: $(AB)C$ or $A(BC)$.
The result will always be the same, but computational costs may vary significantly depending on the matrix dimensions~\cite{reinforcement-learning}.

Similarly, for tensor network contractions, the order in which the tensors are contracted plays an important role.
The cost of the contraction scales, in general, exponentially with the number of tensors~\cite{graph-decomposition}.
When computing a tensor contraction of a given network, a path is used to determine which two tensors to contract at each step.
Finding a good contraction path is essential to improve computation time.

Different approaches have been proposed to compute efficient contraction paths, namely graph decomposition~\cite{graph-decomposition}, 
a modified search algorithm with enhanced pruning~\cite{enhanced-search}, a hypergraph partitioning approach~\cite{hypergraph-partitioning}, methods using machine learning and reinforcement learning~\cite{reinforcement-learning}, simulated annealing and genetic algorithms~\cite{annealing-genetic}, and a greedy algorithm by opt\_einsum~\cite{opt-einsum,opt-einsum-web}.
These methods are explained in more detail in Section~\ref{sec:related-work}.

In this paper, we introduce a novel strategy to improve tensor contraction paths by modifying the current greedy approach.
The standard greedy algorithm~\cite{opt-einsum,opt-einsum-web} applies a cost function to determine the pairwise contractions for the path at each step.
The cost function used is very straightforward and is based on the sizes of the two input tensors and the size of the output tensor.
Our approach focuses on using more information to determine the costs of pairwise contractions.
We provide different cost functions to cover a broad range of problems.
Our approach outperforms the state-of-the-art greedy implementations by opt\_einsum~\cite{opt-einsum,opt-einsum-web,numpy} for a variety of problems, and in some cases even outperforms sophisticated compute-intensive approaches such as hypergraph partitioning combined with greedy~\cite{hypergraph-partitioning,cotengra}.
The experimental results show that our modified greedy algorithm not only computes efficient contraction paths, but also has a better ratio between contraction path quality and optimization time than other path optimizers.
The code to our tool and the experiments can be found on Github: \url{https://github.com/sheela98/Bachelor-Thesis.git}.

This paper is structured as follows: We first provide the necessary background information in Section~\ref{sec:background}, explaining tensors, Einstein summation notation,
tensor hypernetworks, and tensor contraction in more detail. 
We take a look at the current research and approaches to find efficient contraction paths in Section~\ref{sec:related-work}.
In Section~\ref{sec:algorithm}, we give more detailed information about our approach, explaining the three steps of the greedy algorithm and our contribution by providing different cost functions.
We present our benchmarks in Section~\ref{sec:experiments}, comparing our approach to others.
Finally, in Section~\ref{sec:conclusion} we discuss the experimental results, and point out what implications they have for future work.

\section{Background}
\label{sec:background}
In this section we give the necessary background on tensors, Einstein summation notation, tensor hypernetworks, and tensor contraction.

\subsection{Tensors}
Tensors are algebraic objects used in different fields such as computer science, mathematics, and physics.
We define a tensor as a multidimensional array of numbers.
Graphically, we represent a tensor as a node, with the edges representing its indices.

\begin{figure}
    \centering
    \includegraphics{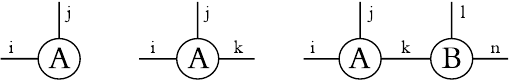}
    \caption{Graphical representation of a matrix, a tensor, and a tensor network.}
\label{fig::tensor-examples}
\end{figure}

Figure~\ref{fig::tensor-examples} illustrates examples, showing a matrix \(A\) with indices \(i,j\), and a tensor \(A\) with indices \(i,j,k\).
The number of indices is referred to as the rank of the tensor. 
Each index has a specific valid range. 
Consider tensor \(A\) with indices \(i,j,k\) and respective ranges \(i \in \{1,2,3,4,5 \}, j \in \{1,2,3\}\) and  \(k \in \{1,2\}\). 
We define the size to be the product of the maximum range of the indices of a tensor. 
The size for tensor \(A\) is computed as follows: \(5 \cdot 3 \cdot 2 = 30\).

\subsection{Einstein Summation Notation}
A tensor contraction is essentially the summation over pairs of repeated indices.
For example, to contract two tensors \(A_{ijk}\) and  \(B_{kln}\) (see Figure~\ref{fig::tensor-examples}),
that share one index \(k\), the contraction is computed by summing over the common index \(k\).
Mathematically, this tensor contraction is the computation of the following expression:
\[ C_{ijln} = \sum_{k} A_{ijk} \cdot B_{kln}.\] 
We can simplify the expression using the Einstein summation notation.
The notation was introduced in 1916 by Einstein and allows to represent tensor expressions succinctly.
The summation is implied over indices that occur more than once.
The same contraction as before can be expressed as:
\[ C_{ijln} = A_{ijk} \cdot B_{kln}.\]
We differentiate between the original notation previously mentioned and modern Einstein notation.
The modern notation is widely used in Machine Learning and Linear Algebra libraries and frameworks such as Optimized Einsum or Numpy~\cite{opt-einsum,opt-einsum-web,numpy}.
The same expression as before can be written as: 
\[ijk, kln \rightarrow ijln.\]
Note that, in modern Einstein notation, the indices of the output tensor are specified after the arrow and all indices that are not included are used for summation.
To simplify naming, we refer to Einstein summation expressions as Einsum expressions in this paper.
We will use both the traditional and modern notation style to represent Einsum expressions, depending on the context.

\subsection{Tensor Hypernetworks}
\begin{figure}[h]
    \centering
    \includegraphics{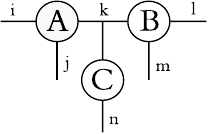}
    \caption{Example of a tensor hypernetwork.}
    \label{fig::hyper-tensor-network}
\end{figure}

An Einsum expression can be represented graphically as a tensor hypernetwork.
Tensor hypernetworks are tensor networks with hyperedges.
A hyperedge occurs when three or more tensors, which are represented as nodes, share at least one index. 
See Figure~\ref{fig::hyper-tensor-network} for an example of a tensor hypernetwork where the index $k$ is shared by tensors $A, B$ and $C$.

\subsection{Tensor Contraction}
If we want to contract a tensor hypernetwork, we have to contract the specified indices.
For example, we have given the following network \(N\):
\[ N = (A \cdot B \cdot C)_{ijlmn} = \sum_{k} A_{ijk} \cdot B_{klm} \cdot C_{jkn}.\]
In original Einstein summation notation, it can be written as:
\[ N = (A \cdot B \cdot C)_{ijlmn} = A_{ijk} \cdot B_{klm} \cdot C_{jkn}.\]
In modern Einstein summation notation, it changes to:
\[ijk,klm,jkn\rightarrow ijlmn.\]
The contraction is usually executed by contracting a pair of tensors at a time. 
For instance, the expression can be computed by first contracting \(A \cdot B\) and then \((AB) \cdot C\) or first \(A \cdot C\) and then \((AC) \cdot B\) or \(B \cdot C\) and then \((BC) \cdot A\).
The order in which each tensor is contracted with another is stated by the contraction path.
All possible combinations lead to the same result, but the execution time can vary drastically. 

Finding an efficient contraction sequence is crucial to improve performance.
Optimized Einsum (opt\_einsum) is a Python package that provides a way to compute tensor contraction paths via various pathfinding algorithms such as a greedy approach, a dynamic algorithm, and an exhaustive search~\cite{opt-einsum,opt-einsum-web}.
Our approach builds on the greedy approach by opt\_einsum.
In Section~\ref{sec:algorithm}, we present our modified greedy algorithm in detail.

\section{Related Work}
\label{sec:related-work} 
Finding efficient contraction paths for tensor hypernetworks has been the focus of several previous works.
Various strategies have been used to solve the ordering problem, from more traditional algorithms using heuristics to machine learning solutions.
 In this section, we introduce the different approaches to compute tensor contraction paths.

For smaller tensor hypernetworks, optimal solutions have been found using a modified search algorithm with enhanced pruning~\cite{enhanced-search}. 
A different approach using a simulated annealing and a genetic algorithm~\cite{annealing-genetic} showed that these methods may outperform the standard greedy approach for smaller networks.

Gray J. investigated hyperparameter optimization by combining various optimization techniques~\cite{hypergraph-partitioning}.
The suggestion was to use graph partitioning combined with various other algorithms to find efficient tensor contraction sequences.

In another work, a graph decomposition~\cite{graph-decomposition} method was explored.
Two methods, Line-Graph (LG) and Factor-Tree (FT), were used for graph decompositions. 
LG is a general purpose method for finding a contraction order using structured graph analysis.
As LG can not handle high-rank tensors, the Factor-Tree (FT) method was used during preprocessing in this scenario.

In a more recent paper from 2022, the problem of finding an efficient path for tensor contractions was formulated as a reinforcement learning (RL) problem~\cite{reinforcement-learning}.
A combination of RL and Graph Neural Networks (GNNs) was used to find an efficient path. 
This approach was especially effective for tensor network problems, simulating quantum circuits. 
The experiments included real and synthetic quantum circuits.

The approach we present in this paper uses a different aspect to improve tensor contraction paths.
We focus on the well-known greedy algorithm~\cite{opt-einsum,opt-einsum-web}.
  We use several cost functions for different problem domains, such as model counting, quantum circuits, graph problems, and language models.
Despite its simplicity, this approach proves to be very effective.
In the next section, we give an in-depth introduction to the greedy algorithm and our modifications to enable our multiple-cost-functions approach.

\section{The Algorithm}
\label{sec:algorithm} 
In this section, we introduce the greedy algorithm.
We give an overview of how the algorithm computes contraction paths, and we show how we can embed our multiple-cost-functions approach into the standard greedy algorithm.

\subsection{The Standard Greedy Algorithm}
The greedy approach was introduced in Optimized Einsum (opt\_einsum)~\cite{opt-einsum,opt-einsum-web} as an efficient strategy to find contraction paths for expressions with large numbers of tensors.
It computes the contraction path in three phases:
\begin{enumerate}
    \item Compute Hadamard products, that is, elementwise multiplications of tensors with the same index sets.
    \item Contract the remaining tensors by choosing at each step the pair with the lowest cost until there are no more contraction indices left.
    \item Compute outer products by choosing at each step the pair that minimizes the sum of the input sizes.
\end{enumerate}
In the standard greedy approach of opt\_einsum, a cost heuristic is used in the second phase to determine which tensor pair to choose for the contraction path.
The second phase is the most cost-intensive part of the algorithm.
To find the pair of tensors to be contracted, opt\_einsum takes the remaining sequence of tensors as input and computes the cost for each possible contraction pair, that is, the output size of the tensor minus the sum of the two sizes of the input tensors.
At each step, the pair with the lowest cost is chosen and added to the path.
The greedy algorithm has space and time complexity of $O(n * k)$~\cite{opt-einsum-web} where $n$ is the number of input tensors and $k$ the maximum number of tensors that share any index to be contracted, excluding indices that occur in every input tensor.
This approach scales well for very large contractions of low-rank tensors where the corresponding hyperedges are small, that is, they connect the tensors sparsely.

\subsection{The Modified Greedy Algorithm}
\paragraph{Modifications.}
Originally, the standard greedy algorithm uses only one cost function in phase two to evaluate the possible pairwise contractions.
We have changed this part of the algorithm to allow custom cost functions to be passed as parameters.
At runtime, we employ a set of different cost functions, each of which is evaluated against the problem instance.
Based on this evaluation, the most suitable cost function is selected and utilized to generate further contraction paths.
The criteria for this selection are determined by the computation's objective, which may include either reducing the size of the largest intermediate tensor or minimizing the total number of floating point operations (flops).

\paragraph{The Cost Functions.} 
An efficient contraction path may vary depending on the problem.
We designed our cost functions by considering problems from different domains, such as model counting, weighted model counting, machine learning, and quantum circuits.
We also used problem instances derived from planar graphs, regular graphs, square grid graphs, and random graphs to tune the cost functions.
Each cost function performs well for a different range of problems.
It is therefore important to consider several cost functions at runtime in order to find the one that works best for a particular problem instance. 
The actual cost functions we developed can be viewed within the code of the algorithm in our repository.

\begin{figure}
    \centering
    \includegraphics{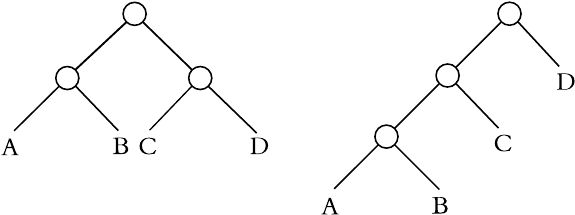}
    \caption{Two example contraction trees for computing the result of a tensor network with four tensors. On the left is a balanced contraction tree, and on the right is a skewed one.}
    \label{fig::sb-contraction-path}
\end{figure}

In the following, we explain the reasons for using multiple cost functions by looking at contraction trees and their influence on total contraction costs.
In Figure~\ref{fig::sb-contraction-path}, there are two different types of tensor contraction trees: one is a balanced, and the other a skewed contraction tree.
Each cost function for computing a contraction path tends to determine the shape of the contraction tree.
For some problems a balanced contraction tree might be the best choice, and for others a skewed one is better.
A balanced contraction tree can commonly be produced by the following cost function: \textit{size12 - min(size1, size2)},
where \textit{size12} is the size of the output tensor, 
and \textit{size1} and \textit{size2} are the respective sizes of the two input tensors for a pairwise contraction.
At each step, we compute the cost for all possible pairs of the remaining tensors that still need to be contracted.
We choose the one with the smallest cost. 
A skewed contraction tree can be made, for example, using the following cost function: \textit{size12 - max(size1, size2)}.

Let us consider an example where we have four tensors $A_i, B_{ij}, C_{jk}, D_{km}$ with dimension sizes $i=3, j=2, k=3$ and $m=2$.
We want to contract the tensors according to the Einsum expression $i, ij, jk, km \rightarrow m$.
If we compute a contraction path with each of the two cost functions, we get two different contraction paths. 
\begin{table}[htbp]
    \centering
    \caption{Contraction paths produced by applying the two example cost functions.}
    \begin{tabularx}{\textwidth}{XX}
        \toprule
        \textbf{size12 - min(size1, size2)} & \textbf{size12 - max(size1, size2)} \\
        \midrule
        $jk, km \rightarrow jm$ & $i, ij \rightarrow j$ \\
        $i, ij \rightarrow j$ & $jk, j \rightarrow k$ \\
        $jm, j \rightarrow m$ & $km, k \rightarrow m$ \\
        \bottomrule
    \end{tabularx}
    \label{tab:table-contraction-path}
\end{table}

The cost function \textit{size12 - min(size1, size2)} results in the contraction path shown on the left in Table~\ref{tab:table-contraction-path}, whereas the cost function \textit{size12 - max(size1, size2)} results in the contraction path shown on the right in Table~\ref{tab:table-contraction-path}.
The resulting contraction path on the left in Table~\ref{tab:table-contraction-path} produces a balanced tree, and the one on the right produces a skewed tree.
The resulting contraction trees correspond to those in Figure~\ref{fig::sb-contraction-path}.
When we consider the operations necessary, to execute the whole Einsum expression, we see that for the balanced tree we need one matrix multiplication, 
and two matrix-vector multiplications. 
For the skewed contraction tree, we only need three matrix-vector multiplications.   
To execute the whole Einsum expression, we need 44 flops with the balanced contraction tree, and 36 flops with the skewed one.

As we can see, there is a significant difference in performance for the two cost functions.
This is why different problems require different cost functions to find an efficient contraction path.
Our multiple-cost-functions approach specifically targets this issue, finding efficient tensor contraction paths for a broad range of problems.

\section{Experiments}
\label{sec:experiments}
In this section, we first give an overview of the setup we use to evaluate the modified greedy algorithm.
We then look at the results of the experiments. 
Finally, we interpret the results.
\subsection{Setup}
To evaluate our approach, we apply our modified greedy algorithm to 40 problems covering a wide range of domains, including model counting, weighted model counting, machine learning and quantum circuits.
In addition, we also benchmark problem instances derived from planar graphs, regular graphs, square grid graphs, and random graphs.
In Table~\ref{tab:table-problems}, the benchmark problems are divided into four different categories.
Furthermore, Table~\ref{tab:table-problems} contains the number of instances per category and the respective minimum and maximum number of tensors that occur for a problem instance in the respective category.
\begin{table}[htbp]
  \centering
  \caption{Problem instances used in the experiments.}
  \begin{tabularx}{\textwidth}{Xlll}
    \toprule
    \textbf{Category} & \textbf{Instance Count} & \textbf{Min Tensors} & \textbf{Max Tensors} \\
    \midrule
    Model Counting & 10 & 353 & 10699 \\
    Quantum Circuits & 10 & 28 & 3206 \\
    Graph Problems & 10 & 60 & 781 \\
    Language Models & 10 & 38 & 84 \\
    \bottomrule
  \end{tabularx}
  \label{tab:table-problems}
\end{table}

We compare our implementation with the greedy and random greedy approach used in opt\_einsum~\cite{opt-einsum,opt-einsum-web,numpy}, 
the greedy algorithm employed in cotengra~\cite{hypergraph-partitioning,cotengra}, 
and the hypergraph partitioning approach also used by cotengra. 
The current state of the art is the latter approach, which relies heavily on the hypergraph partitioner KaHyPar~\cite{kahypar}.
The experiments are performed on a MacBook Pro with an Apple M1 8-core processor running Sonoma 14.2.1 with 8 GB of RAM.
To compile the modified algorithm, we use g++13. 
The evaluation is done in Python 3.12.

\subsection{Results}
To evaluate our multiple-cost-functions approach, we compute contraction paths for ten problems from each of the four categories.
We compare different algorithms and measure the flops for each algorithm.
We perform two experiments.
In the first experiment, we compute for each problem instance 128 paths with each algorithm.
The first experiment aims to evaluate the quality of the solution without considering the computation time.
In the second experiment, we perform the same computation, but instead of limiting the number of paths, we limited the computation time to one second.
The second experiment aims to demonstrate the trade-off between time and quality, because in practical scenarios an efficient path should be computed quickly so that the computation of the whole Einsum expression can start without much delay.
We repeat each experiment five times and report the median value for each algorithm and the corresponding problem instance.

\begin{figure}[htbp]
  \centering
	\includegraphics[width=0.90\textwidth]{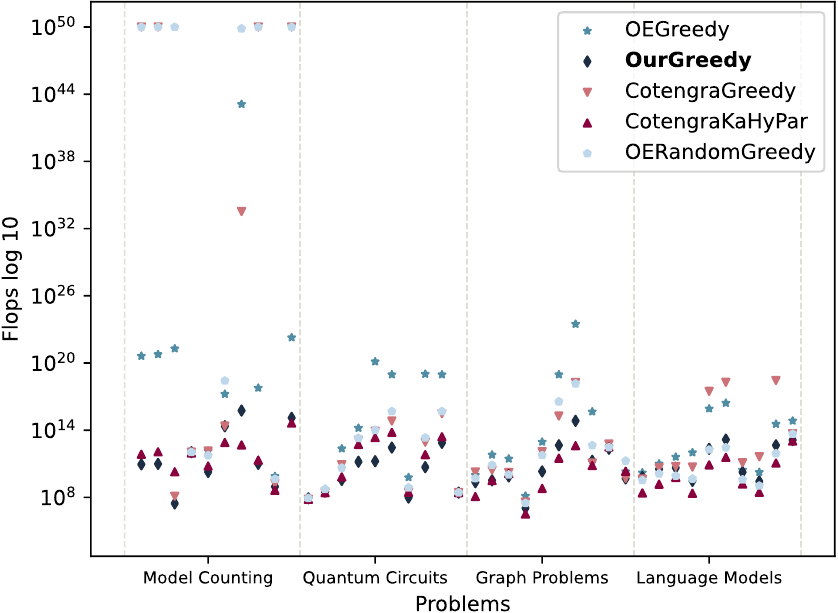}
  \caption{Results for the computation of 128 paths.}
  \label{fig::results-128-paths}
\end{figure}

\begin{figure}[htbp]
    \centering
		\includegraphics[width=0.90\textwidth]{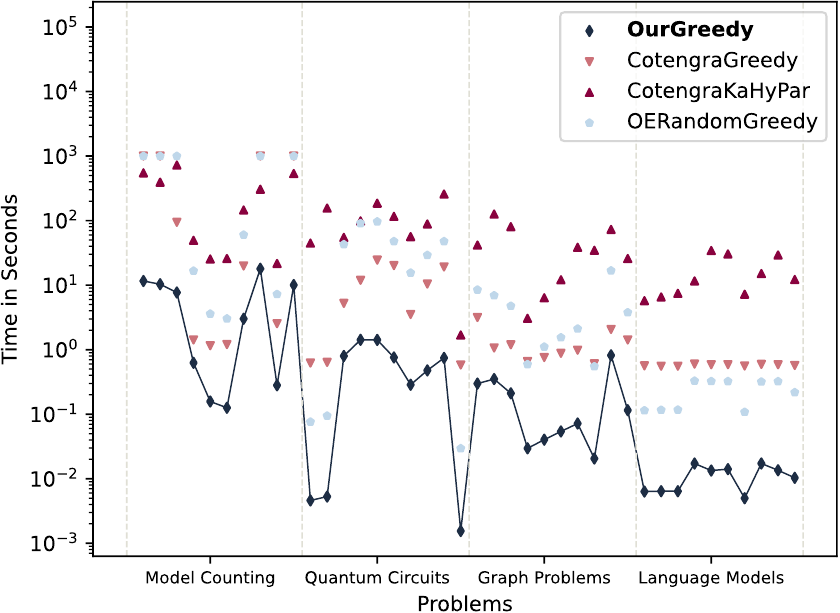}
  \caption{Time for the computation of 128 paths.}
	\label{fig::results-128-paths-time}
\end{figure}

\begin{figure}[tbp]
  \centering
	\includegraphics[width=0.90\textwidth]{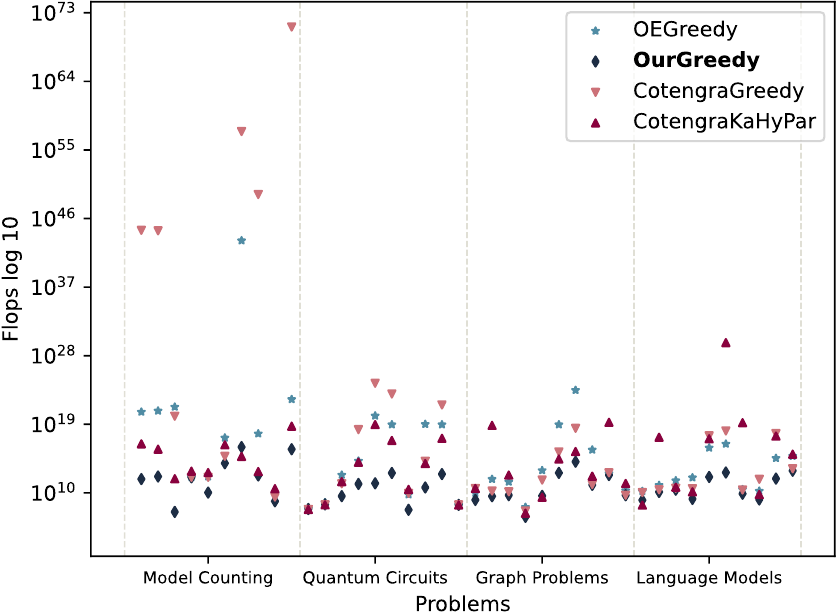}
  \caption{Results with timeout of one second.}
  \label{fig::results-timeout}
\end{figure}

Figure~\ref{fig::results-128-paths} and Figure~\ref{fig::results-timeout} show the graphical representations of the experimental results. 
The x-axis represents the problem categories (model counting, quantum circuits, graph problems, and language models), and the y-axis represents the flops on a logarithmic scale.
Figure~\ref{fig::results-128-paths} indicates that our approach outperforms others in certain categories, such as quantum circuits. 
However, for graph problems, cotengra's hypergraph partitioner stands out, 
and for language models, the random greedy approach by opt\_einsum~\cite{opt-einsum,opt-einsum-web,numpy} and the hypergraph partitioner~\cite{hypergraph-partitioning,cotengra} seem to perform slightly better. 
It is important to note that the first experiment has no time limit.
Also note that the random greedy approaches of opt\_einsum and cotengra failed to compute contraction paths for some large model counting problems.
These failed problem instances are shown in Figure~\ref{fig::results-128-paths} on $y=10^{50}$ flops.

Although the first experiment has no time limit, we show in Figure~\ref{fig::results-128-paths-time} the actual runtimes for computing 128 paths for each problem instance from Figure~\ref{fig::results-128-paths}.
The problem instances that could not be executed are shown at $y=10^3$.
It is clear that our approach significantly outperforms all other algorithms in terms of performance when considering the time to compute 128 paths.
Therefore, our approach is less resource hungry compared to other approaches.
We are able to compute more paths in a given time frame and therefore potentially find more efficient paths in short time spans.
To substantiate our claim, we repeat the computation from Figure~\ref{fig::results-128-paths} with a timeout of one second instead of limiting it to 128 paths.
The results are shown in Figure~\ref{fig::results-timeout}.

Introducing a time limit of one second clearly influences the results. 
Figure~\ref{fig::results-timeout} shows that our multiple-cost-functions approach achieves better results than all other algorithms in all problem categories.
The reason for this observation are twofold: first, we can compute more paths in a given time frame, and second, the different cost functions yield strongly varying contraction trees, which quickly leads to the selection of a suitable cost function for a problem instance.
Overall, our approach quickly converges to an efficient contraction path, which is not necessarily the best possible contraction path, but still has a satisfactory number of flops.

\section{Conclusions}
\label{sec:conclusion}
We have introduced a multiple-cost-functions approach to compute contraction paths for tensor hypernetworks.
At runtime, each cost function is evaluated against the problem instance.
The best cost function is chosen and used to compute the contraction path.
To evaluate our approach we conducted several experiments comparing our multiple-cost-functions algorithm to commonly used algorithms such as the standard greedy and random greedy algorithm by opt\_einsum~\cite{opt-einsum,opt-einsum-web,numpy}, 
and the greedy algorithm and hypergraph partitioning approach employed in cotengra~\cite{hypergraph-partitioning,cotengra}.
We found that our approach can compute efficient contraction paths in less time and solve large problems where other approaches fail.
In the future, we plan to extend the presented multiple-cost-functions approach to compute efficient contraction paths for even larger problems.

 \FloatBarrier

\bibliographystyle{unsrt}  
\bibliography{references}

\begin{thebibliography}{10}

\bibitem{reinforcement-learning}
Eli Meirom et~al.
\newblock Optimizing tensor network contraction using reinforcement learning.
\newblock In {\em International Conference on Machine Learning}, 2022.

\bibitem{hypergraph-partitioning}
Johnnie Gray and Stefanos Kourtis.
\newblock Hyper-optimized tensor network contraction.
\newblock {\em Quantum}, 2021.

\bibitem{many-body}
Pietro Silvi, Ferdinand Tschirsich, et~al.
\newblock The tensor networks anthology: Simulation techniques for many-body
  quantum lattice systems.
\newblock {\em SciPost Physics Lecture Notes}, 2019.

\bibitem{graph-decomposition}
Jeffrey~M. Dudek, Leonardo Due{\~{n}}as{-}Osorio, and Moshe~Y. Vardi.
\newblock Efficient contraction of large tensor networks for weighted model
  counting through graph decompositions.
\newblock {\em arXiv: 1908.04381}, 2020.

\bibitem{enhanced-search}
Robert~NC Pfeifer, Jutho Haegeman, and Frank Verstraete.
\newblock Faster identification of optimal contraction sequences for tensor
  networks.
\newblock In {\em Physical Review E}, 2014.

\bibitem{annealing-genetic}
Frank Schindler and Adam~S. Jermyn.
\newblock Algorithms for tensor network contraction ordering.
\newblock {\em Machine Learning: Science and Technology}, 2020.

\bibitem{opt-einsum}
Daniel~G. Smith and Johnnie Gray.
\newblock Opt\_einsum - a python package for optimizing contraction order for
  einsum-like expressions.
\newblock {\em Journal of Open Source Software}, 2018.

\bibitem{opt-einsum-web}
Daniel~G. Smith and Johnnie Gray.
\newblock Optimized einsum.
\newblock \url{https://dgasmith.github.io/opt_einsum/}.
\newblock (retrieved: 07.02.2024).

\bibitem{numpy}
NumPy Developers.
\newblock Numpy - einsum.
\newblock
  \url{https://numpy.org/doc/stable/reference/generated/numpy.einsum.html}.
\newblock (retrieved: 07.02.2024).

\bibitem{cotengra}
Johnnie Gray.
\newblock Cotengra.
\newblock \url{https://github.com/jcmgray/cotengra}.
\newblock (retrieved: 07.03.2024).

\bibitem{kahypar}
Sebastian Schlag, Tobias Heuer, Lars Gottesbüren, Yaroslav Akhremtsev,
  Christian Schulz, and Peter Sanders.
\newblock High-quality hypergraph partitioning.
\newblock {\em ACM Journal of Experimental Algorithmics}, 2022.

\end{thebibliography}

\end{document}